\documentclass[twocolumn,showpacs,floatfix,prb,superscriptaddress,eqsecnum]{revtex4}
\usepackage{graphicx}
\usepackage{amsmath}
\usepackage{bm}

\begin{document}
\title{Bi-stability in voltage-biased NISIN structures}
\author{I. Snyman}
\affiliation{National Institute for Theoretical Physics, Private Bag X1, 7602 Matieland, South Africa}
\affiliation{Instituut-Lorentz, Universiteit Leiden, P.O. Box 9506, 2300 RA Leiden, The Netherlands}
\author{Yu. V. Nazarov}
\affiliation{Kavli Institute of Nanoscience, Delft University of Technology, 2628 CJ Delft, The Netherlands} 
\date{August 2008}
\begin{abstract}
As a generic example of a voltage-driven superconducting structure
we study a short superconductor 
connected to normal leads by means of low transparency
tunnel junctions, with a voltage bias $V$ between the leads.
The superconducting order parameter $\Delta$ is to be determined self-consistently.
We study the stationary states of the system as well as 
the dynamics after a perturbation. 
We find a region in parameter space 
where there are two stable stationary states at a given voltage. 
These bi-stable states are distinguished by distinct
values of the superconducting order parameter 
$\Delta$ and of the current between the leads.
We have evaluated 
(1) the multi-valued superconducting order parameter $\Delta$ at given $V$; 
(2) the current between
the leads at a given V; and (3) 
the critical voltage at which 
superconductivity in the island ceases.
With regards to dynamics, we find numerical evidence
that the stationary states are stable and that no complicated
non-stationary regime can be induced by changing the voltage.
This result is somewhat unexpected
and by no means trivial, given the fact that the
system is driven out of equilibrium. The response to a change in the
voltage is always gradual, even in the regime
where changing the interaction strength induces rapid anharmonic oscillations
of the order parameter.
\end{abstract}
\pacs{74.40.+k, 74.78.Fk, 74.25.Fy, 74.78.Na \hfill NITheP-08-08}

\maketitle
\section{Introduction}
Electron transport devices combining 
superconducting (S), insulating (I) and normal metal (N) elements
are known as superconducting hetero-structures. Often such 
hetero-structures 
are more than the sum of their parts.\cite{Lam98,Bee95} Phenomena that are not
present in bulk S, I or N systems appear when a device contains
junction between these components. The following examples
are well known: (1) The conductance of a high transparency NS junction
does not equal the conductance of the normal metal on its own, as one might naively
expect. If the normal metal is free of impurities, the conductance is {\em higher}
than that of the normal metal.\cite{Blo82}  
This surprising effect is due to a process known as Andreev reflection.\cite{And64} 
During Andreev reflection at an NS interface,
an electron impinging on the interface from the N side is
reflected back as a hole, while a Cooper pair propagates away
from the interface on the S side. 
(2) In Josephson junctions, the simplest of which is perhaps the SIS 
hetero-structure,\cite{Jos62}
a DC current can flow at zero bias voltage. This happens when the
superconducting phase difference across the junction is non-zero.\cite{Tin96}

The above examples can be understood in terms of equilibrium properties
of the hetero-structure. When a superconducting device is perturbed outside
equilibrium, yet more interesting effects can occur,\cite{Kopnin}
for instance, oscillations under stationary non-equilibrium conditions.
An elementary example: 
if a Josephson junction is biased with a DC (i.e. fixed) voltage, 
an AC (i.e. oscillating) current flows through the junction.\cite{Tin96} 
Another example of the kind has been investigated
in the context of cold Fermi gases 
in optical traps. In these systems, the interaction between atoms can be tuned and changed
by means of a so-called Feshbach resonance. If the interaction is attractive,
the gas forms a BCS-condensate. 
Recent studies\cite{Bar06,Yuz05} have considered what happens
if the value of the attractive pairing interaction is changed abruptly.
It was discovered that, depending on the ratio between the initial and final 
values of the interaction strength, the condensate order parameter
can perform anharmonic oscillations that do not decay in time.

The initial motivation for the research presented in this paper came 
form the study of Keizer {\em et al.},\cite{Kei06} where the authors investigated
the suppression of the superconducting order parameter by a
voltage applied to a superconducting wire. It was assumed
that $\Delta$ remains stationary. However, this assumption does
not seem well-justified: the stationary voltage could induce
periodic oscillations of $|\Delta|$ or even richer chaotic dynamics. 
Thus prompted, we wanted to address the validity of this assumption
for a decidedly simpler NISIN structure, namely 
a short superconductor
connected to normal leads by means of tunnel junctions.
The structure is biased with a voltage $V$. 

We require that (1) the dominant energy relaxation mechanism in the superconductor
is the tunneling of electrons to the leads, and
(2) spatial variations of the superconducting order parameter
inside the superconductor are negligible. 
To meet the first requirement, the superconductor must have dimensions
smaller than the inelastic scattering length of quasi-particles.
This is not an unrealistic requirement given current
experimental techniques. To meet the second requirement, the
superconductor should firstly contain impurities or have an irregular shape, 
so that the electron wave-functions of the isolated island are isotropic on the
scale of the superconducting coherence length.\cite{deG66}
Secondly, the tunnel junctions connecting it to the leads 
should have a bigger normal-state resistance than that of the superconductor proper. 
In this case, opening up the system by
connecting leads does not re-introduce spatial anisotropy 
of wave-functions inside the island.

The study of NISIN structures has a long history.\cite{Giaver,Hidaku}
Our study complements several previous studies.\cite{Kei06,San01,Mar95} These
dealt with quasi-one-dimensional
superconducting wires between normal leads. Setups 
where either the superconductor was impurity-free or the transparency of
the NS interfaces were high were considered. For these
setups, spatial variations of the order parameter, 
specifically the spatial gradient of the superconducting phase, 
can be large. Including these spatial variations in the description
of the superconductor significantly complicates matters. Hence these
studies focused on numerical calculations and assumed that the 
superconducting order-parameter and all other quantities of interest
were stationary. It should also be mentioned that asymmetric 
couplings, where the superconductor is
coupled more strongly to one lead than the other, did not receive 
detailed analysis. The only asymmetric setup considered consisted
of one interface with tunable transparency and the other perfectly
transparent.\cite{Mar95} One of the main conclusions of these studies is
that, if the bias voltage is large enough, the system switches to the
normal state. Some evidence for
a bi-stable region where, depending on the history of the system,
either the superconducting or the normal
state can occur at a given voltage, was reported.\cite{Kei06}

The absence of spatial variations in the 
system we study allows us to perform analytical calculations, 
provided we assume stationarity.
Results are obtained for an arbitrary ratio of the coupling strengths
to the leads. We derive
transcendental equations relating the superconducting order parameter 
to the bias voltage, and derive an explicit formula for the 
current between the leads.  
As mentioned, the assumption of stationarity is
however not a priori justified. As was seen in the examples mentioned
at the beginning of this introduction, 
non-equilibrium conditions in superconductors 
often go hand in hand with non-stationary  
behavior of observable quantities. Indeed, the NISIN junction
that we study is a non-linear system subjected to a driving force (and
to damping). Non-linearity here means that the dynamical equations
for one-particle Green functions are not linear in the Green
functions. This is due to the existence of a non-zero
superconducting order parameter. 
The driving force is provided by the voltage
(and the damping by tunneling of electrons from the island into the leads).
Non-linear driven systems (think of the nonlinear pendulum) often have
chaotic dynamics. The assumption of stationarity would miss this.
We therefore supplement
our analytical calculation with numerical calculations that study
the dynamics in real time. 

Our main results are the following: The stationary states that we found
analytically are stable. 
Furthermore, there is a parameter region where two different
stationary states are stable at the same voltage. 
(This is the ``bi-stability'' of the title.)
For a symmetric coupling to the source and drain leads,
one of the two states is superconducting (characterized by a non-zero
order parameter) and the other is normal. 
Since we are in the regime of high tunnel barriers, at a given voltage,
the superconducting
island allows less current to flow between the leads than the island
in the normal state.\cite{Blo82}
This current is a directly measurable quantity and 
allows one to distinguish between superconducting and normal states.
For some asymmetric couplings however, 
both the stable states are superconducting. We have calculated the
current that flows between the leads at a given voltage, and at arbitrary asymmetry
of the coupling to the two leads. 
We find that the value of the current also
allows one to distinguish between different stable superconducting states
at a given voltage. 

The time-dependent calculations revealed that 
once the bias voltage becomes constant in time, the system always
relaxes into one of the stationary states. Non-stationary
behavior of physical quantities always decays in time, unlike in
the case of a DC-biased Josephson junction. (Despite it being a non-linear system, 
a superconductor driven by a voltage is therefore fundamentally 
different from a nonlinear pendulum driven by an external force.)
If the bias voltage
is changed slowly, an initial stationary state evolves adiabatically.
By changing the voltage slowly 
we have observed the expected hysteresis associated with the 
existence of two stable states at some voltages. 

The rest of the paper is structured as follows. 
In Sec.~\ref{sec:model} we specify the model to be studied, and
present the equations that determine its state. In Sec.~\ref{sec:stat}
we solve these equations analytically,
assuming that the system is in a stationary state.
We analyze the stationary stationary states we find and calculate
the $I$-$V$ characteristic of the system. In Sec.~\ref{sec:dyn}
we establish that the stationary states are the only stable
states of the DC-biased system. We do so by studying the dynamics
of the system after a perturbation. In Sec.~\ref{sec:con}
we summarize our main results.    

\section{Model}
\label{sec:model}
As stated in the introduction,
we consider a superconducting island connected to two normal 
leads by means of low transparency tunnel barriers.
The superconducting order parameter is taken to be spatially
isotropic inside the island. The physical requirements for this
condition to hold have already been discussed in the introduction. 
We assume that the dominant energy relaxation mechanism for
the superconductor is tunneling of electrons to the leads.
For given barrier transparencies, this restricts the size of
the superconductor to less than the inelastic scattering length
of quasi-particles inside the superconductor.

Our analysis of the system is based on the Keldysh Green function
technique.\cite{Kel64,Ram86,Brink03} We start our discussion of the equations
governing the system by defining the necessary Green functions. 
 
\subsection{Definition of Green functions}
The Green functions are expectation values of
products of the 
Heisenberg operators $a_{m\pm}^\dagger(t)$ and $a_{m\pm}(t)$ that create and annihilate electrons
in levels of the isolated island.
Here $m$ labels single particle levels. The $\pm$ index accounts for 
Kramer's degeneracy. 
As we are dealing with a problem
involving superconductivity, all Green functions are $2\times 2$ matrices in Nambu space. 
It is useful to define Nambu space matrices $\eta_j$, $j=0,\ldots,3$ 
such that $\eta_0$ is the identity matrix and $\eta_1$, $\eta_2$ and $\eta_3$ are the standard Pauli matrices.
We also define matrices $\eta_\pm=(\eta_1\pm i\eta_2)/2$.

The retarded ($R$),
Keldysh ($K$) and advanced ($A$) Green functions of each level are defined as\cite{Note1}
\begin{widetext}
\begin{subequations}
\begin{eqnarray}
R_m(t,t')&=&-i\eta_3\left<\left(\begin{array}{rr}\{a_{m+}(t),a_{m+}^\dagger(t')\}&\{a_{m+}(t),a_{m-}(t')\}\\
\{a_{m-}^\dagger(t),a_{m+}^\dagger(t')\}&\{a_{m-}^\dagger(t),a_{m-}(t')\}\end{array}\right)\right>\theta(t-t'), \label{eq:retarded_def}\\
K_m(t,t')&=&-i\eta_3\left<\left(\begin{array}{rr}{[a_{m+}(t),a_{m+}^\dagger(t')]}&{[a_{m+}(t),a_{m-}(t')]}\\
{[ a_{m-}^\dagger(t) , a_{m+}^\dagger(t') ]} & {[ a_{m-}^\dagger(t) , a_{m-}(t') ]}\end{array}\right)\right>,\label{eq:keldysh_def}\\
A_m(t,t')&=&\eta_3 R_m(t',t)^\dagger\eta_3.\label{eq:advanced_def} 
\end{eqnarray}
\end{subequations}
\end{widetext}

The Green functions are grouped into a matrix
\begin{equation}
G_m(t,t')=\left(\begin{array}{cc}R_m(t,t')&K_m(t,t')\\0&A_m(t,t')\end{array}\right).\label{eq:green_def}
\end{equation}
This further $2\times 2$ matrix structure is referred to as Keldysh space.
As with Nambu space, it is useful to define matrices $\tau_j$, $j=0,\ldots,3$.
The matrix $\tau_j$ is the same as the matrix $\eta_j$ but now operating in Keldysh space.
We also carry over the definition of $\tau_\pm$ from Nambu space. 
A basis for the $4\times4$ matrices that result from combining Keldysh and Nambu indices 
is constructed by means of a tensor product
$\tau_j\otimes\eta_k$, with the $\tau$'s always acting in Keldysh space
and the $\eta$'s in Nambu space.

The quantities that we calculate, namely the order parameter $\Delta(t)$ and the current $I(t)$,
are collective in the sense that they result from the sum of the contributions of all the
individual levels. 
Accordingly a formalism exists that does not require knowledge of the 
Green functions of individual levels
but only the sums\cite{Ram86,Bel99,Naz05,Naz99,Naz94}
\begin{equation}
{\cal G}(t,t')=\frac{i\delta_s}{\pi}\sum_m {\cal G}_m(t,t'),\hspace{2mm}{\cal G}=G,\,R,\,K,\,A,
\label{eq:totalg}
\end{equation} 
that are known as quasi-classical Green functions.
Here $\delta_s$ is the mean level spacing of the island. 

We will work with the quasi-classical Green functions 
throughout the present section. The advantage of doing so is that the theory can be formulated
with the least amount of clutter. When doing time-dependent numerics in Sec.~\ref{sec:dyn}
however, we find it more convenient to work with the Green functions of the individual levels.
In principle though, the theory outlined in this section, following as it does from the theory outlined in 
Sec.~\ref{sec:dyn}, gives exactly the same answers.    

\subsection{Equations of motion}
\label{sec:eqsofm}
The equations that determine the Green functions can be derived from
the circuit theory of non-equilibrium superconductivity.\cite{Naz05,Naz99,Naz94}
Viewed as a matrix in time, Nambu and Keldysh indices,
the Green function $G$ satisfies the commutation relation\cite{Note2}
\begin{equation}
\left[H-\Sigma,G\right]=0.\label{eq:dyson_g}
\end{equation}
Here $H$ describes the dynamics of the isolated superconductor:
\begin{subequations}
\label{statH}
\begin{eqnarray}
H(t,t')&=&\tau_0\otimes\eta_3\,\delta(t-t')\left[i\partial_t-h(t)\right],\label{eq:zero_order}\\
h(t)&=&\left(\begin{array}{rr}-\mu_s(t)&\Delta(t),\\
\Delta(t)^*&\mu_s(t)\end{array}\right).\label{eq:BdG_Ham}
\end{eqnarray}
\end{subequations}
The matrix $h(t)$ is a remnant of the Bogoliubov-de Gennes Hamiltonian.\cite{deG66}
Bearing in mind that we consider a non-equilibrium setup, we must allow
the order parameter $\Delta(t)$ and the chemical potential $\mu_s(t)$ of the superconductor to be time-dependent.
Their values at each instant in time are determined by imposing self-consistency. 

The time derivative standing to the right of $G$ in the term $GH$ of 
Eq.~(\ref{eq:dyson_g}) can be shifted to act on the second time argument of $G$
at the cost of a minus sign, i.e.
\begin{eqnarray}
\int d\tilde{t}\, G(t,\tilde{t})\partial_{\tilde{t}}\delta(\tilde{t}-t')&=&
-\int d\tilde{t}\, \partial_{\tilde{t}}G(t,\tilde{t})\delta(\tilde{t}-t')\nonumber\\
&=&-\partial_{t'}G(t,t').
\end{eqnarray}

The self-energy contains a term corresponding to each lead, i.e. 
\begin{equation}
\Sigma=\Sigma^{(l)}+\Sigma^{(r)},
\end{equation}
$l$ and $r$ referring to the left and right leads respectively. 
The leads act as reservoirs, broadening the island levels to a
finite lifetime and determining their filling.
The self-energy of lead $j$ is $\Sigma^{(j)}=-i\Gamma_j G^{(j)}$,
where Green function $G^{(j)}$ of lead $j$ is defined similarly to the Green 
function of the superconductor (Eq.~\ref{eq:totalg}),
with the sum now running over states in the lead. 
Here $\Gamma_j$ is the tunneling rate from any island level to lead $j$. (For simplicity, we take
the rates associated with different levels to be the same.) 
The leads are large compared to the superconductor, and therefore
$G_j$ does not depend on the state of the superconductor. Furthermore, since the leads
are normal, the off-diagonal Nambu space matrix elements of the lead Green functions are zero.
Explicitly then,
the Green function for lead $j=l,\,r$ has the form
\begin{equation}
G^{(j)}(t,t')=\left(\begin{array}{cc}R^{(j)}(t,t')&K^{(j)}(t,t')\\0&A^{(j)}(t,t')\end{array}\right),\label{eq:res_def}
\end{equation}
with
\begin{subequations}
\begin{align}
R^{(j)}(t,t')&=\delta(t-t')\eta_3=-A^{(j)}(t,t'),\label{eq:res_RA_def}\\
K^{(j)}(t,t')&=2\left(\begin{array}{cc}\sigma_j(t,t')&0\\0&\sigma_j(t,t')^*\end{array}\right).\label{eq:Sigma_K_def}
\end{align}
\end{subequations}
The function $\sigma_j$ describes the distribution of particles in lead $j$.
In general it is given by 
\begin{equation}
\sigma_{j}(t,t')=\int\frac{dE}{2\pi}e^{-iE(t-t')}[1-2f_j(E)]e^{-i\left[\phi_j(t)-\phi_j(t')\right]},\label{eq:sigma_j_def}
\end{equation}
where $f_j(E)$ is the filling factor of states at energy $E$ in lead $j$.
The phase $\phi_j$ sets the time-dependent chemical potential $\mu_j(t)=\partial_t \phi_j(t)$ in lead $j$.
The time-dependent bias voltage between the leads is
\begin{equation}
V(t)=(\mu_l(t)-\mu_r(t))/e,
\label{eq:v}
\end{equation}
where $e$ is the electron charge.
It is convenient to define the total inverse lifetime or Thouless energy $E_{\rm Th}=\Gamma_l+\Gamma_r$ and a
dimensionless symmetry parameter $\gamma=(\Gamma_l-\Gamma_r)/E_{\rm Th}$. For a perfectly symmetric
coupling to the leads, $\gamma=0$ while $\gamma=\pm1$ corresponds to the island being coupled to only
one of the two leads.

The commutator equation (\ref{eq:dyson_g}) on its own is not enough to specify $G$ uniquely. Indeed
what Eq. (\ref{eq:dyson_g}) says is that $G$ has the same eigenstates as $H-\Sigma$, but it does not say anything
about the eigenvalues of $G$. Additional to Eq. (\ref{eq:dyson_g}) there is a also relation between the eigenvalues of $G$ 
and those of $H-\Sigma$.\cite{She85} Let $\left|\lambda\right>$ be a simultaneous eigenstate of $H-\Sigma$ and $G$, such
that its eigenvalue with respect to $H-\Sigma$ is $\lambda$. Then its eigenvalue with respect to $G$ is
${\rm sgn}\left({\rm Im}(\lambda)\right)$. (One can show that the eigenvalues of $H-\Sigma$ come in complex
conjugate pairs and that there are no purely real eigenvalues.) Hence $G$
squares to unity, i.e.
\begin{equation}
G^2=I.
\label{eq_Gsquare}
\end{equation} 

\subsection{Gauge invariance}
At this point we have defined three different Fermi-energies, namely that of the superconductor $\mu_s(t)$ and
those of the leads $\mu_j(t)$, $j=l,\,r$. Since the reference point
from which energy is measured is arbitrary, there is some redundancy. This redundancy is encoded in
a symmetry of the equations for the Green function and boils down to gauge-invariance. 
Consider a transformation on the Green function
\begin{subequations}
\begin{align}
G&\to\tilde G=UGU^\dagger,\\
U(t,t')&=\delta(t-t')\,\tau_0\otimes\exp(i\eta_3\Lambda(t)).
\end{align}
\end{subequations}
As is easy to verify, $\tilde G$ obeys equations of the same form as $G$, with chemical potentials and 
the order parameter transformed according to
\begin{subequations}
\begin{eqnarray}
\mu_j(t)&\to&\tilde{\mu}_j(t)=\mu_j(t)+\partial_t \Lambda(t),\,j=s,\,l,\,r,\nonumber\\
\\
\Delta(t)&\to&\tilde{\Delta}(t)=\Delta(t)\exp[2i\Lambda(t)].\label{eq:deltagauge}
\end{eqnarray}
\end{subequations}
When considering stationary solutions we will fix the gauge by demanding that 
$\Delta$ is time-independent. When considering non-stationary solutions we will
fix the gauge such that the reference point from which chemical potentials are measured is
halfway between the chemical potentials of the reservoirs, i.e. $\mu_{r(l)}(t)=+(-)eV(t)/2$.

\subsection{Self-consistency of $\Delta$}
The value of the order parameter is set by the self-consistency condition
\begin{eqnarray}
\Delta(t)&=&g\delta_s\sum_m \left<a_{m-}(t)a_{m+}(t)\right>\nonumber\\
&=&-\frac{\pi g}{2} {\rm Tr}\left[\eta_-K(t,t)\right],\label{eq:selfcondelta}
\end{eqnarray}
where $g>0$ is the dimensionless pairing interaction strength.
This self-consistency equation suffers from the usual logarithmic divergence which
requires regularization by introducing a large energy cut-off $E_{\rm c.o.}$.
We define $\Delta_0$ as the order parameter of
an isolated superconductor at zero temperature for given $g$ and $E_{\rm c.o.}$.
\begin{equation}
\Delta_0=\frac{E_{\rm c.o.}}{\sinh\tfrac{1}{g}}\Rightarrow\frac{1}{g}=\int_0^{E_{\rm c.o.}}
\frac{dE}{\sqrt{E^2+\Delta_0^2}}.
\end{equation}
This definition then allows us to express $\Delta$ in Eq.~(\ref{eq:selfcondelta})
in terms of $\Delta_0$ rather than in terms of $E_{\rm c.o.}$ and $g$. 

\subsection{Current and chemical potential}
The current from the superconductor into reservoir $j$ is\cite{Note3}
\begin{widetext}
\begin{equation}
I_j(t)=\frac{\pi}{2e}G_j\int dt'\,{\rm Tr}\Big[\tau_-\otimes\eta_3\Big(G(t,t')G^{(j)}(t',t)
-G^{(j)}(t,t')G(t',t)\Big)\Big],\hspace{2mm}G_{l(r)}=(1+(-)\gamma)\frac{E_{\rm Th}}{\delta_s}\frac{e^2}{[\hbar]}.
\label{eq:currentj}
\end{equation}
\end{widetext}
Here $G_j$ is the tunneling conductance of the tunnel barrier between lead $j$ and the superconductor, and
we have indicated in square brackets a factor of $\hbar$ which equals unity in the units we use throughout the paper.  
The total rate of change of the charge in the superconductor equals minus the sum of the currents
to the leads, i.e.
\begin{equation}
-\frac{d}{dt}Q(t)=I_l(t)+I_r(t).
\end{equation}
The charge in the superconductor is related to the chemical potential $\mu_s$ 
by means of the capacitance $C$ of the superconductor, so that $\mu_s$ has to obey
\begin{equation}
\frac{1}{e}\frac{d}{dt}\mu_s(t)=C\frac{d}{dt}Q(t).\label{eq:dmudt}
\end{equation} 
When the system is not stationary, this equation sets the value of $\mu_s(t)$ at each instant in time,
since $dQ(t)/dt$ can be calculated directly from $G(t,t')$. 

\subsection{Summary}
In summary then, our task is to find the Green function $G$ as defined in Eq.~(\ref{eq:totalg}) of the
superconductor. In general, the procedure for doing this is as follows: We make an Ansatz for 
the order parameter $\Delta(t)$ and the chemical potential $\mu_s(t)$. We then diagonalize
the operator $H-\Sigma$ (that depends on $\Delta$ and $\mu$). The Green function $G$ is
constructed in the eigenbasis of $H-\Sigma$, according the prescription of Sec.~\ref{sec:eqsofm}.
Subsequently we judge the correctness of the Ansatz for $\Delta(t)$ and $\mu(t)$ by
inquiring whether Eqs.~(\ref{eq:selfcondelta}) and (\ref{eq:dmudt}) are satisfied.

\section{Stationary solutions}
\label{sec:stat}
We consider a time-independent bias voltage between the left and right reservoirs.
In this case the chemical potentials $\mu_l$ and $\mu_r$ of the reservoirs are time-independent. 
We make the Ansatz that the chemical potential $\mu_s$ and the order parameter $\Delta$ of the
superconductor are also
time-independent. The Green function $G(t,t')$ only depends on the time-difference $t-t'$.
It is convenient to
work with the Fourier transformed Green function $G(E)$ which is related to $G(t,t')$ by
\begin{equation}
G(t,t')=\int\frac{dE}{2\pi}e^{-iE(t-t')}G(E).
\end{equation}
It is also convenient to construct a traceless operator
$M=H-\Sigma-\tau_0\otimes\eta_0\,\mu_s$ with Keldysh structure
\begin{equation}
M=\left(\begin{array}{cc}M_R&M_K\\0&M_A\end{array}\right)\label{eq:m_def}.
\end{equation}
In the energy representation the components of $M$ have the explicit form
\begin{subequations}
\begin{eqnarray}
M_R(E)&=&\left(\begin{array}{cc}E+iE_{\rm Th}&-\Delta\\\Delta^*& -E-iE_{\rm Th}\end{array}\right),\\
M_A(E)&=&\left(\begin{array}{cc}E-iE_{\rm Th}&-\Delta\\\Delta^*& -E+iE_{\rm Th}\end{array}\right),\\
M_K(E)&=&2iE_{\rm Th}\left(\begin{array}{cc}\sigma(E)&0\\0&\sigma(-E)\end{array}\right),\\
\sigma(E)&=&\frac{1-\gamma}{2}\sigma_l(E)+\frac{1+\gamma}{2}\sigma_r(E).
\end{eqnarray}
\end{subequations}
We take the left and right leads to be in local zero-temperature equilibrium at Fermi energies
$\mu_l=\mu+eV/2$ and $\mu_r=\mu-eV/2$ so that the filling factors in both reservoirs is a step function
$f_j(E)=\theta(-E)$ and from Eq.~(\ref{eq:sigma_j_def}) follows
\begin{subequations}
\begin{eqnarray}
\sigma_l(E)&=&{\rm sgn}\left(E-\mu-eV/2\right),\\
\sigma_r(E)&=&{\rm sgn}\left(E-\mu+eV/2\right),
\end{eqnarray}
\end{subequations}
where $\mu$ is the average chemical potential $(\mu_r+\mu_l)/2$ in the leads, in the gauge where the
phase of the order parameter is time-independent. The value of $\mu$ will later
be determined by requiring self-consistency of the order parameter $\Delta$.
The Green function $G(E)$ obeys $[M(E),G(E)]=0$. The retarded, advanced and Keldysh components
of this equation are
\begin{subequations}
\begin{align}
&[M_R(E),R(E)]=[M_A(E),A(E)]=0,\label{eq_ret_ad}\\
&M_R(E)K(E)+M_K(E)A(E)\nonumber\\
&\hspace{5mm}-R(E)M_K(E)-K(E)M_A(E)=0\label{eq_kel}.
\end{align}
\end{subequations}
\begin{figure}[th]
\begin{center}
\includegraphics[width=.95 \columnwidth]{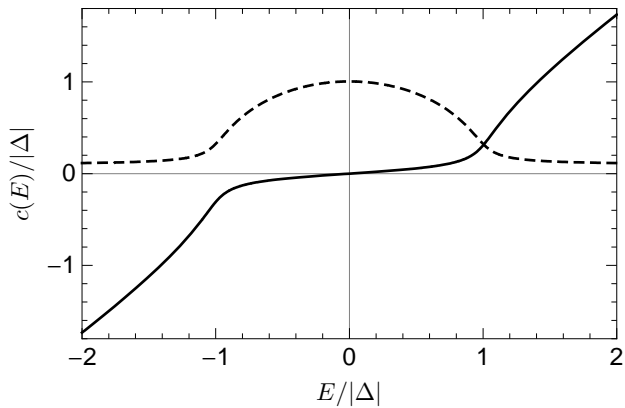}
\caption{The function $c(E)$ as defined in Eq.~(\ref{eq:c}) frequently appears in expressions associated with
stationary solutions. The solid line represents the real part and the dashed line the imaginary part.
The Thouless energy was taken as $E_{\rm Th}=0.1\,|\Delta|$.\label{fig:c}}
\end{center}
\end{figure}
With the aide of the prescription below Eq.~(\ref{eq:v}) for choosing the eigenvalues of $G$, one then readily
finds for the retarded and advanced Green functions
\begin{subequations} 
\label{eq:rasols}
\begin{eqnarray}
R(E)&=&\frac{1}{c(E)}\left(\begin{array}{cc}E+iE_{\rm Th}&-\Delta\\\Delta^*& -E-iE_{\rm Th}\end{array}\right),\\
A(E)&=&\frac{1}{c(-E)}\left(\begin{array}{cc}E-iE_{\rm Th}&-\Delta\\\Delta^*& -E+iE_{\rm Th}\end{array}\right),\\
c(E)&=&\sqrt{(E+iE_{\rm Th})^2-|\Delta|^2}.\label{eq:c}
\end{eqnarray}
\end{subequations}
The function $c(E)$, which we will frequently encounter, is defined with branch cuts along the lines
$E_\pm=\pm|\Delta|\pm x-iE_{\rm Th}$ with $x$ real and positive. The branch with $\lim_{E\in R\to\pm\infty}c(E)/E=1$
is taken. Considered as a function of real $E$, the real part of $c(E)$ is odd, and the imaginary part
is even and positive so that
\begin{equation}
c(E)^*=-c(-E).
\end{equation}
The real and imaginary parts of $c(E)$ is plotted for real $E$ in Fig.~\ref{fig:c}.

\begin{figure}[tbh]
\begin{center}
\includegraphics[width=.95 \columnwidth]{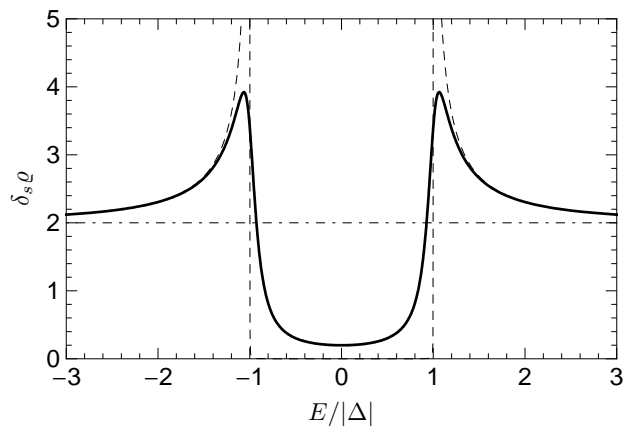}
\caption{The density of states of the superconducting island (Eq. \ref{eq:dos}) for finite Thouless energy (solid line).
The dashed line shows the density of states of the isolated superconducting island with the same $|\Delta|$, while the
horizontal dot-dashed line shows the density of states of the normal island. 
A value of $E_{\rm Th}=0.1\,|\Delta|$ was used.\label{fig:dos}}
\end{center}
\end{figure}

Note that $R(E)^2=A(E)^2=\eta_0$ as required by Eq.~\ref{eq_Gsquare}.
In general the superconducting density of states is
$\varrho(E)={\rm Tr}\,\eta_3\left[R(E)-A(E)\right]/2\delta_s$
so that we find from the solutions for $R$ and $A$ (Eq.~\ref{eq:rasols})
\begin{equation}
\varrho(E)=\frac{2}{\delta_s}{\rm Re}\left[\frac{E+iE_{\rm Th}}{c(E)}\right]\label{eq:dos}.
\end{equation}
The density of states for an isolated superconductor has singularities at energies $E=\pm|\Delta|$
of the form $1/\sqrt{E^2-|\Delta|^2}$. The coupling to the leads regularizes the singularities at
an energy scale of $E_{\rm Th}$. Furthermore, whereas the density of states of the isolated superconductor
vanishes for energies $|E|<|\Delta|$, the coupling to the leads softens the gap so that there are some
states for energies $|E|<|\Delta|$ as shown in Fig.~\ref{fig:dos}.  

The next step is to solve Eq.~(\ref{eq_kel}) for $K(E)$. Here we use the fact
that $R(E)K(E)A(E)=-K(E)$, which follows from the requirement that $G^2=I$ (Eq.~\ref{eq_Gsquare}).
Note also that $R(E)=M_R(E)/c(E)$ and $A(E)=M_A(E)/c(-E)$. Using these identities and
multiplying Eq.~(\ref{eq_kel}) from the left by $R(E)$, we find
\begin{equation}
K(E)=\frac{1}{c(E)+c(-E)}\left[M_K(E)-R(E)M_K(E)A(E)\right].
\end{equation}
After some algebra we obtain 
\begin{equation}
K(E)=\left(\begin{array}{cc}K^{(1)}(E)&K^{(2)}(E)\\-K^{(2)}(E)^*&K^{(1)}(-E)\end{array}\right),
\end{equation}
where 
\begin{widetext}
\begin{subequations}
\begin{eqnarray}
K^{(1)}(E)&=&\delta_s\varrho(E)\sigma(E)
-\frac{|\Delta|^2}{E}{\rm Re}\left[\frac{1}{c(E)}\right]\left[\sigma(E)+\sigma(-E)\right],\\
K^{(2)}(E)&=&-\Delta{\rm Re}\left[\frac{1}{c(E)}\right]\left\{
\left[\sigma(E)-\sigma(-E)\right]-\frac{iE_{\rm Th}}{E}\left[\sigma(E)+\sigma(-E)\right]\right\}.
\end{eqnarray}
\end{subequations}
\end{widetext}
Having obtained $K(E)$ we can find $|\Delta|$ and $\mu$ from the self-consistency condition 
Eq.~(\ref{eq:selfcondelta}). Below we write the real and imaginary parts of the 
self-consistency equation separately. The real part reads
\begin{equation}
0=\int dE\left\{{\rm Re}\left[\frac{1}{c(E)}\right]\frac{\sigma(E)-\sigma(-E)}{2}
-\frac{1}{\sqrt{E^2+\Delta_0^2}}\right\},
\end{equation}
while the imaginary part reads
\begin{equation}
0=\int dE\,\frac{1}{E}{\rm Re}\left[\frac{1}{c(E)}\right]\left[\sigma(E)+\sigma(-E)\right].\label{selfconim}
\end{equation}
These integrals can be done explicitly. We use the identities
\begin{subequations}
\begin{align}
&\int_0^EdE'\,{\rm Re}\frac{1}{c(E')}=F_R(E)-F_R(0),\\
&\int_0^EdE'\,\frac{1}{E'}{\rm Re}\frac{1}{c(E')}=\frac{1}{\sqrt{E_{\rm Th}^2+|\Delta|^2}}F_I(E),\\
\end{align}
\end{subequations}
where
\begin{subequations}
\begin{eqnarray}
F_R(E)&=&\ln\left|\frac{E+iE_{\rm Th}+c(E)}{\Delta_0}\right|,\\
F_I(E)&=&\arctan\left[\frac{{\rm Re}[c(E)]}{\sqrt{E_{\rm Th}^2+|\Delta|^2}}\right].
\end{eqnarray}
\end{subequations}
Here the branch for which $-\pi/2<\arctan(x)<\pi/2$ is implied.
Thus we obtain the transcendental equations
\begin{subequations}
\label{eq:transcendental}
\begin{align}
&0=(1-\gamma)F_R(\tfrac{eV}{2}+\mu)+(1+\gamma)F_R(\mu-\tfrac{eV}{2})\label{transcendentala},\\
&0=(1-\gamma)F_I(\tfrac{eV}{2}+\mu)+(1+\gamma)F_I(\mu-\tfrac{eV}{2})\label{transcendentalb},
\end{align}
\end{subequations}
that determine $|\Delta|$ and $\mu$ for given $V$, $E_{\rm Th}$ and $\gamma$.
Below we solve these equations analytically in certain limiting cases and 
numerically for more general cases. Only the amplitude of $\Delta$ is fixed
by these equations. By choosing the appropriate gauge [cf. Eq.~(\ref{eq:deltagauge})], 
we can set the phase of $\Delta$ to any value. In the rest of this section we therefore
drop the absolute value notation, and take $\Delta$ real and positive.

Before explicitly finding $\Delta$ and $\mu$, we calculate the current from 
Eq.~(\ref{eq:currentj}) and the solution for $K(E)$. Assuming that 
the self-consistency equation (Eq.~\ref{selfconim}) is fulfilled, we find that
the current $I_r$ from the superconductor to the right lead equals minus the
current $I_l$ from the superconductor to the left lead, as it should.
For the current $I=I_r=-I_l$ from the left lead to the right lead we find
\begin{eqnarray}
I&=&\frac{(1-\gamma^2)e E_{\rm Th}}{2}\int_{\mu-\tfrac{eV}{2}}^{\mu+\tfrac{eV}{2}}dE\,\varrho(E)\nonumber\\
&=&\frac{G_N}{e}{\rm Re}\left[c(\mu+\tfrac{eV}{2})-c(\mu-\tfrac{eV}{2})\right].
\label{eq:current}
\end{eqnarray}
Here $G_N$ is the series conductance of the tunneling barriers to the leads 
\begin{equation}
\label{eq:GN}
G_N=\left[G_l^{-1}+G_r^{-1}\right]^{-1},
\end{equation}
and $G_l$ and $G_r$ are the junction conductances given in Eq.~(\ref{eq:currentj}).
 
\begin{figure}[tbh]
\begin{center}
\includegraphics[width=.95 \columnwidth]{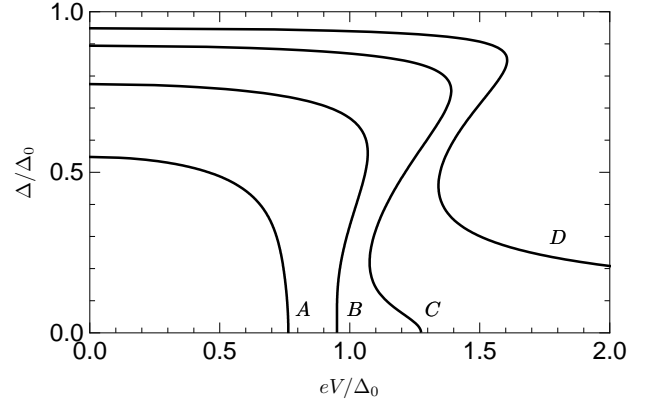}
\caption{
The order parameter $\Delta$ versus voltage $V$, for given $E_{\rm Th}$
and $\gamma$. 
Curves $A$, $B$, $C$ and $D$ respectively correspond to $E_{\rm Th}=0.35\Delta_0$ and $\gamma=0.2$;
$E_{\rm Th}=0.2 \Delta_0$ and $\gamma=0.075$; 
$E_{\rm Th}=0.1\Delta_0$ and $\gamma=0.1$; and 
$E_{\rm Th}=0.01\Delta_0$ and $\gamma=0.3$. \label{fig:selfcond}}
\end{center}
\end{figure}

\begin{figure}[tbh]
\begin{center}
\includegraphics[width=.95 \columnwidth]{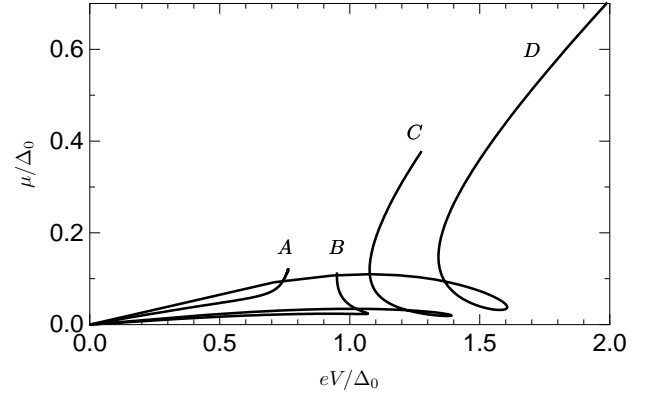}
\caption{The chemical potential $\mu$ versus voltage $V$, for given $E_{\rm Th}$ and $\gamma$. 
Curves $A$, $B$, $C$ and $D$ correspond to the same parameter values as in Fig.~\ref{fig:selfcond}.
\label{fig:selfconm}}
\end{center}
\end{figure}

Now we investigate the transcendental equations
(Eqs. \ref{eq:transcendental}) for $\mu$ and $\Delta$. There are
three parameters, namely $E_{\rm Th}$, $\gamma$ and $V$ that determine the solution. Two of
these, $E_{\rm Th}$ and $\gamma$, are fixed for a given device, while the voltage $V$ can 
be varied for a given device. (Recall that $E_{\rm Th}$ measures the overall coupling to the leads, while
$\gamma$ measures the degree of asymmetry between the two lead couplings.) 
Hence it is natural to specify values for $E_{\rm Th}$ and $\gamma$ and then
consider $\Delta$, $\mu$ and $I$ as functions of $V$. In Fig.~\ref{fig:selfcond}
we show four curves of $\Delta$ versus $V$, each corresponding to a different choice
of the parameters $E_{\rm Th}$ and $\gamma$. In Fig.~\ref{fig:selfconm} we show the
corresponding curves of $\mu$ versus $V$.  

Let us firstly note the general trend that increasing
$E_{\rm Th}$ leads to a smaller order parameter. The reason for this is that 
$E_{\rm Th}^{-1}$ is the typical time an electron remains in the superconductor.
The shorter this time (the larger $E_{\rm Th}$) the harder it is for electrons
to form Cooper pairs, and superconductivity is inhibited.
Secondly, note that at large enough $E_{\rm Th}$ the order parameter is a decreasing function
of $V$. We can therefore obtain the critical Thouless energy $E_{\rm Th}^{(c)}$ beyond which superconductivity
vanishes by setting $V$ to zero and asking how large can we make $E_{\rm Th}$ before $\Delta$ becomes zero.

In the case of $V=0$, the self-consistency equations are solved by $\mu=0$ and
\begin{equation}
\Delta=\left\{\begin{array}{lr}\Delta_0\sqrt{1-\frac{2E_{\rm Th}}{\Delta_0}}&E_{\rm Th}<\Delta_0/2\\
0&E_{\rm Th}>\Delta_0/2\end{array}\right. .
\end{equation}
From this we conclude that the critical Thouless energy is $E_{\rm Th}^{(c)}=\Delta_0/2$.

Having established the range of $E_{\rm Th}$ in which superconductivity persists, we 
now take a closer look at $\Delta$ as a function of $V$. We have chosen the parameters of the
four solutions in Fig.~\ref{fig:selfcond} to show all the different possible shapes
that curve of $\Delta$ versus $V$ can take. We see that at a given voltage $V$ there can be either 
zero, one, two or three non-zero solutions $\Delta$. 

To characterize the different types of curve, we consider $V$ as a function of $\Delta$ on
the interval $\Delta\in[0,\Delta_0\sqrt{1-2E_{\rm Th}/\Delta_0}]$.
In curves of the type $A$ in Fig.~\ref{fig:selfcond}, $V$ is a monotonically decreasing function
of $\Delta$. In contrast, curves of type $B$, $C$ and $D$ have local extrema. A curve of type
$B$ has a local minimum at the left boundary $\Delta=0$ of the $\Delta$ interval on which the
function $V(\Delta)$ is defined. Then the curve reaches a maximum at some intermediate value $\Delta_1$,
before dropping to zero at the right boundary $\Delta=\Delta_0\sqrt{1-2E_{\rm Th}/\Delta_0}$.
Curves $C$ and $D$ are distinguished from curve $B$ by the fact that $V$ reaches a local maximum
instead of a minimum at the left boundary $\Delta=0$ of the $\Delta$ interval. 
There is another local maximum at intermediate $\Delta_1$ before $V$ drops to zero at
$\Delta=\Delta_0\sqrt{1-2E_{\rm Th}/\Delta_0}$. In curves of type $C$, the absolute maximum of
$V$ as a function of $\Delta$ is at the intermediate value $\Delta_1$ while for
curves of type $D$ the absolute maximum of $V$ is at $\Delta=0$. 

\begin{figure}[tbh]
\begin{center}
\includegraphics[width=.95 \columnwidth]{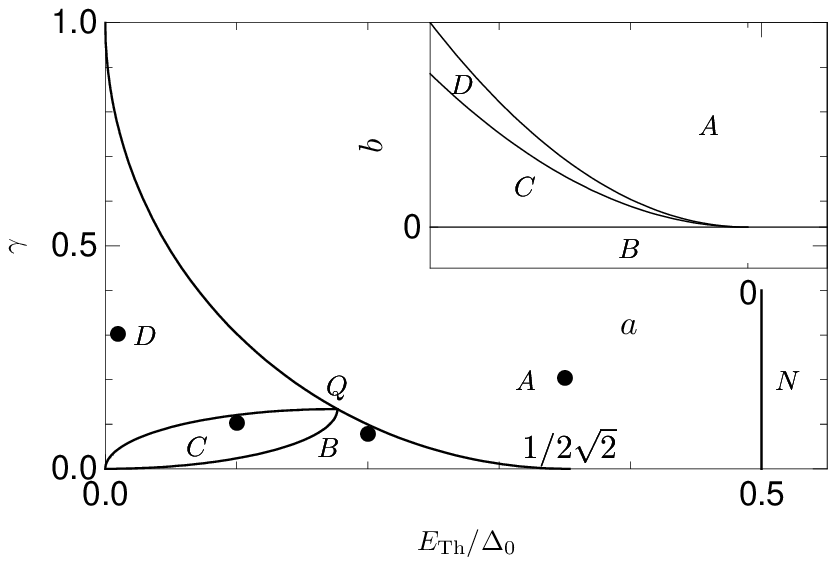}
\caption{Schematic diagram of the partitioning of the $E_{\rm Th}$---$\gamma$ parameter
space into regions where the curve of $\Delta$ versus $V$ is of the types $A$, $B$,
$C$ and $D$ (Fig.~\ref{fig:selfcond}).
The regions $A$, $B$, $C$, and $D$ meet at point $Q$. The line $E_{\rm Th}=\Delta_0/2$ separates
the normal and superconducting regions of parameter space. The dots
in the figure indicate the parameter values that correspond to the curves in
Fig.~\ref{fig:selfcond}. The inset shows the regions $A$, $B$, $C$, and $D$ in the
parameter space $a$---$b$ of the polynomial $\tilde{V}(\Delta)$ of Eq.~(\ref{eq:pol}).
The topology of the $E_{\rm Th}$---$\gamma$ parameter space of the superconductor in the region
of the point $Q$ can be understood by considering the topology of the parameter space of the polynomial. 
\label{fig:parspace}}
\end{center}
\end{figure}

Next we ask how the $E_{\rm Th}$---$\gamma$ parameter space is divided into regions $A$, $B$, $C$ and $D$
corresponding to the respective types of solution of the self-consistency equations. Specifically,
which regions share a mutual border? Assuming that the function $V(\Delta)$ changes smoothly as 
$E_{\rm Th}$ and $\gamma$ are varied, the transitions 
$A\leftrightarrow B$, $B\leftrightarrow C$, $C\leftrightarrow D$ and
$D\leftrightarrow A$ are possible. The transition $A\leftrightarrow C$ is not possible. Whenever one tries to
smoothly deform a curve of type $A$ in Fig.~\ref{fig:selfcond} to a curve of type $C$, one
invariably reaches a curve of type $B$ or $D$ during an intermediate stage of the deformation.
Similarly the transition $B\leftrightarrow D$ is impossible. A smooth deformation of a curve of type
$B$ into one of type $D$ passes through an intermediate stage where the curve is of types $A$ or $C$.
To illustrate these ideas we consider a polynomial equation of the form
\begin{equation}
\frac{\tilde V}{\Delta_0}=\frac{V_0}{\Delta_0}-\frac{1}{6}\left(\frac{\Delta}{\Delta_0}\right)^6
-\frac{a}{4}\left(\frac{\Delta}{\Delta_0}\right)^4
-\frac{b}{2}\left(\frac{\Delta}{\Delta_0}\right)^2.
\label{eq:pol}
\end{equation}
We ask what are the respective regions of the $a$---$b$ plane in which $\tilde V(\Delta)$ is a curve of 
type $A$, $B$, $C$ and $D$. Region $A$, where 
$\tilde{V}(\Delta)$ is of type $A$, is given by $a>0,\,b>0$ or $a<0,\,b>a^2/4$. Region $B$ where
$\tilde{V}(\Delta)$ is of type $B$ consists of all points $(a,b)$ such that $b<0$. Region $C$
consists of all points $(a,b)$ such that $a<0$ and $0<b<3a^2/16$. Region $D$ consists of all points
$(a,b)$ such that $a<0$ and $3a^2/16<b<a^2/4$. The regions and their borders are shown in 
the inset in Fig.~\ref{fig:parspace}.
The most pertinent feature of the figure is that the four
distinct regions meet in the single point $a=b=0$.
 
Based on a combination of numerical and analytical results we have concluded that the 
$E_{\rm Th}$---$\gamma$ parameter space has a very similar topology to this polynomial example. 
(In principle it could have differed from the polynomial example by having disconnected regions of the same type,
for instance two islands of region $D$, one embedded in a sea of region $A$, the other in a sea 
of region $C$.) Fig.~\ref{fig:parspace} is a schematic diagram of how the $E_{\rm Th}$---$\gamma$
parameter space is partitioned into regions $A$, $B$, $C$ and $D$. The following features of the
diagram are conjectures based on numerical evidence: 
(1) The regions of types $A$, $B$, $C$ and $D$ are simply connected.
(2) The border between regions $A$ and $D$ starts at the corner
$\gamma=1$, $E_{\rm Th}=0$.
Other features are deduced from analytical results:
(1) The line $\gamma=0$, $E_{\rm Th}<\Delta_0/2\sqrt{2}$ belongs to region $B$. 
(2) The line $\gamma=0$, $\Delta_0/2\sqrt{2}<E_{\rm Th}<\Delta_0/2$ belongs to region $A$.
(3) For $E_{\rm Th}>\Delta_0/2$ the system is in the normal state while it is superconducting for
$E_{\rm Th}<\Delta_0/2$.
4) The border of regions $D$ and $C$ meets the border of region $B$ and $C$ at $E_{\rm Th}=0$, 
$\gamma=0$.  

In region $D$, superconductivity can persist up to voltages that are large compared to 
$\Delta_0$. For given $E_{\rm Th}$ and $\gamma$ there is however always a critical voltage $V_c$
beyond which superconductivity ceases. (This is a second order phase-transition.)
For the voltage $V_c$ we have obtained the following
analytical result from Eq. (\ref{eq:transcendental}). At finite $\gamma$ and for $E_{\rm Th}$ sufficiently small, 
$V_c$ obeys the power law 
\begin{equation}
V^{(c)}=\frac{\Delta_0}{2e}\left[\frac{2E_{\rm Th}}{\Delta_0}\sec\frac{\pi\lambda}{2}\right]^{-\frac{1}{\lambda}},
\hspace{2mm}\lambda=\frac{1-|\gamma|}{1+|\gamma|}.
\label{eq:power law}
\end{equation}
This power law is valid as long as $V^{(c)}\gg\Delta_0/e$.
It is from this result that we are able to conclude that the region of finite $\gamma$ and infinitesimal $E_{\rm Th}$
belongs to region $D$.

\begin{figure}[tbh]
\begin{center}
\includegraphics[width=.95 \columnwidth]{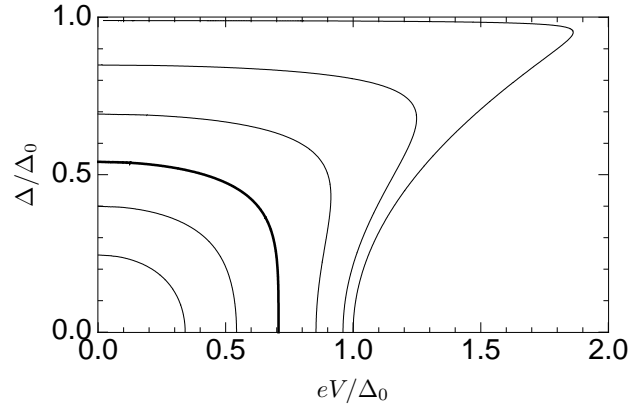}
\caption{The order parameter $\Delta$ versus voltage $V$ for symmetric coupling to the leads, i.e. $\gamma=0$,
according to Eq.~(\ref{eq:g0gap}). Different curves correspond to different $E_{\rm Th}$. From the top curve 
to bottom curve we took $E_{\rm Th}/\Delta_0=.01$, $0.14$, $0.26$, $1/2\sqrt{2}(\simeq0.35)$, $0.42$ and $0.47$.
The curve corresponding to $E_{\rm Th}=1/2\sqrt{2}\Delta_0$ is plotted thicker than the others. For smaller 
$E_{\rm Th}$ are of type $B$ with two non-zero values for $\Delta$ at some voltages. For larger $E_{\rm Th}$,
curves are of type $A$, with at most one non-zero $\Delta$ at every voltage.  
\label{figg0gaps}}
\end{center}
\end{figure}

Another analytical result can be obtained for the case of perfectly symmetric coupling to the leads, i.e. $\gamma=0$.
In this case, Eq. (\ref{transcendentalb}) is solved by $\mu=0$ and the relation between $\Delta$ and $V$ can be stated as
\begin{equation}
eV=\Delta_0\left(1+\frac{\Delta^2}{\Delta_0^2}\right)
\sqrt{1-\frac{4E_{\rm Th}^2/\Delta_0^2}{\left(1-\frac{\Delta^2}{\Delta_0^2}\right)^2}}.
\label{eq:g0gap}
\end{equation}
This result is plotted for several values of $E_{\rm Th}$ in Fig.~\ref{figg0gaps}.
It is from this result that we are able to conclude that the line segment $\gamma=0$, $0<E_{\rm Th}<\Delta_0/2\sqrt{2}$
belongs to region $B$ while the line segment $\gamma=0$, $\Delta_0/2\sqrt{2}<E_{\rm Th}<\Delta_0/2$ belongs to
region $A$. The $E_{\rm Th}\to0$ limit of Eq.~(\ref{eq:g0gap}) can be obtained by considering a bulk superconductor
and assuming a quasi-particle distribution function $n(E)=(\theta(-eV/2-E)+\theta(eV/2-E))/2$. It is also worth
noting that the same result is obtained for a T-junction where the stem of the 
T is a superconductor and the bar a voltage-biased
dirty normal metal wire.\cite{Kei06}

\begin{figure}[tbh]
\begin{center}
\includegraphics[width=.95 \columnwidth]{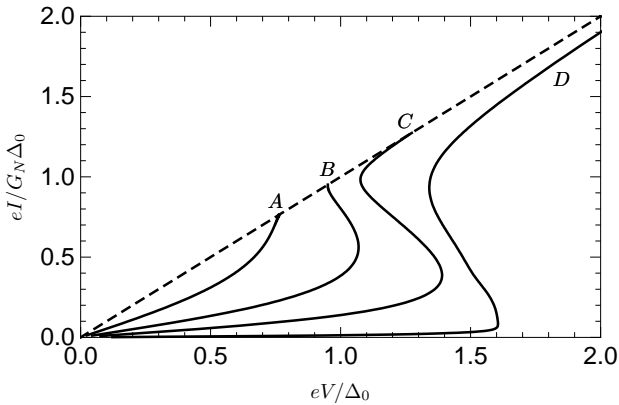}
\caption{The current $I$ through the superconductor versus the voltage $V$ across it. Curves $A$,
$B$, $C$ and $D$ correspond to the respective parameter values quoted in 
Figs.~\ref{fig:selfcond} and \ref{fig:selfconm}. The dashed line shows the current through the system in
the absence of superconductivity.\label{fig:current}}
\end{center}
\end{figure}

Finally, we consider the $I-V$ curves associated with the solutions $\Delta$ and $\mu$ of Figs.~\ref{fig:selfcond} and
\ref{fig:selfconm}. The results are shown in Fig.~\ref{fig:current}. From these curves we can infer the results
that will be obtained in an experiment in which the voltage $V$ is swept adiabatically from zero to
several $\Delta_0/e$ and back to zero. In region $A$ of parameter space there is a single current
associated with each voltage. At some voltage $V_+$ of order $\Delta_0/e$ the
system makes a phase transition to the normal state, but this does not
lead to a discontinuity in the current versus voltage curve.
In contrast, in regions $B$, $C$ and $D$, 
the current will make discontinuous jumps as the voltage is swept. Hysteresis will also be observed. Suppose the
device is in region $B$ of parameter space. As $V$ is swept from $0$ upwards, a voltage $V_+$
is crossed where the current makes a finite jump. After the jump, the system is in the normal state and
the current is $G_N V$. ($G_N$ is the normal state conductance of the setup, cf. Eq.~\ref{eq:GN}.) 
On the backward sweep from $V>V_+$ to zero, the system remains normal when $V_+$ is reached. At
some voltage $V_-<V_+$ the current jumps from its value $G_NV_-$ in the normal state to a smaller value,
signaling the onset of superconductivity. The behavior of the system in region $C$ of parameter space is
similar. The upward sweep of the voltage produces a jump in the current at a voltage $V_+$. After the
jump the system is normal and the current is given by $I=G_N V$. The difference from region $B$ appears
when the voltage is swept back from $V_+$ to zero. At some voltage smaller than $V_+$ the current 
starts deviating from its value in the normal state, but there is no discontinuous jump yet. Even so,
the system has turned superconducting. When the jump in current now occurs at $V_-<V_+$, the system
switches between two different superconducting states. Finally, for parameters in region $D$, the
voltage sweep produces results similar to that in region $C$. The difference between regions $C$ and 
$D$ is that in $D$ the system also jumps between two superconducting states at $V_+$ during the forward
sweep. 

\section{Dynamics}
\label{sec:dyn}
We concluded the previous section with a discussion of hysteresis in the 
current--voltage characteristic of the superconducting island. The conclusions we drew
rely on the assumption that after the system is perturbed by a change in the
bias voltage, it relaxes into a stationary state. 
The validity of this assumption is by no means obvious,
since the system is driven (by the bias voltage) and the stationary state 
is not an equilibrium state. Frankly, our own initial expectation was that
the presence of a bias voltage would cause
the dynamics of $|\Delta(t)|$ to be quasi-periodic or
chaotic. We therefore did numerical simulations in order to investigate
the dynamics of $|\Delta(t)|$ in the presence of a bias voltage.
Our main result is this:
Suppose the bias voltage assumes the constant value $V_f$ for times $t>t_f$.
Then (contrary to our original expectations) at $t\gg t_f$ the superconductor 
will always be found in one of the 
stationary states associated with $V_f$. This is true regardless of the history of the
system prior to $t<t_f$. In particular, the time dependence of the bias voltage prior to $t_f$
does not matter. Nor does the state of the superconductor prior to $t_f$ matter.
Only when there is more than one non-zero stationary solution associated with $V_f$
does the history of the system have {\em any} baring on its final state. 
In this case, the history of the system determines which of the possible stationary
states eventually becomes the final state of the superconductor. 
For slowly varying voltages, the predictions of the previous section regarding hysteresis 
are confirmed. 
In this section we discuss the
numerics that yielded the above results.

For the purpose of numerics we find it 
advantageous not to take the sum over levels of the Green function
as we did in the previous sections. Instead we work with the Green functions
of each individual level. The advantage of this scheme is that it allows
us to work with ordinary differential equations. From these differential
equations it is straight-forward to construct a time-series in
which the next element can be calculated if the present elements are known.
As far as we can see, no such `local in time' update equations exist for
the Green functions summed over levels.
Naturally there are disadvantages to working with the individual level Green functions as well;
the number of equations to be solved numerically is increased significantly. 
As a result the calculation is computationally expensive and therefore time-consuming.

The Green functions of the individual levels obey the equations
\begin{equation}
(H_m-\Sigma)G_m=G_m(H_m-\Sigma)=I.\label{Dyn:Dyson}
\end{equation}
Here $H_m$ differs from the operator $H$ that appeared in Eq.~(\ref{statH}) in that it
contains the energy $\varepsilon_m$ of level $m$. It is explicitly given by
\begin{subequations}
\label{Dyn:H}
\begin{eqnarray}
H_m(t,t')&=&\tau_0\otimes\eta_3\,\delta(t-t')\left[i\partial_t-h_m(t)\right],\\
h_m(t)&=&\left(\begin{array}{cc}\varepsilon_m-\mu_s(t)&\Delta(t)\\
\Delta(t)^*&\mu_s(t)-\varepsilon_m\end{array}\right).
\end{eqnarray}
\end{subequations}
The operator $h_m(t)$ is the time-dependent Bogoliubov-de Gennes Hamiltonian.\cite{deG66}
The self-energy $\Sigma$ is the same as in Sect.~\ref{sec:eqsofm}.

We measure energies from a point halfway between the chemical
potentials of the leads. As a result the phases $\phi_{j}(t)$
that appear in the reservoir self-energies are 
$\phi_{r(l)}(t)=+(-)\phi(t)/2$ where $\phi$ is related
to the voltage $V$ by $V(t)=\partial_t\phi(t)/e$.

We parameterize the Green functions in terms of a set
of auxiliary functions. This 
eliminates some redundancies that are present due to symmetries 
of the equations of motion. We start by noting that since the retarded
and advanced Green functions are related by Eq.~(\ref{eq:advanced_def}),
we do not need to consider both. We work with the retarded Green function.
We define a matrix $r_m(t,\tau)$ that is related to $R_m(t,t-\tau)$ by
the equation
\begin{subequations}
\begin{eqnarray}
R_m(t,t-\tau)&=&i\,\eta_3~r_m(t,\tau),\\
r_m(t,\tau)&=&r_m^{(0)}(t,\tau)\,\eta_0+i\,\bm{r}_m(t,\tau)\cdot\bm{\eta}.\label{Dyn:r}
\end{eqnarray}
\end{subequations}
Here the component $r_m^{(0)}(t,\tau)$ of $r_m(t,\tau)$ is a scalar
function whereas the other three components are grouped into a vector 
$\bm{r}_m(t,\tau)$ such that
\begin{equation}
\bm{r}_m(t,\tau)=\left(r^{(1)}_m(t,\tau),r^{(2)}_m(t,\tau),r^{(3)}_m(t,\tau)\right).
\end{equation}
The vector $\bm{\eta}=(\eta_1,\eta_2,\eta_3)$ contains the Pauli matrices in
Nambu space. Before the voltage bias between the leads is established,
(i.e. for $t\leq0$ and all $\tau$), the functions $r_m^{(0)}(t,\tau)$
and $\bm{r}_m(t,\tau)$ are real.
When the equations of motion (\ref{Dyn:Dyson})
for the retarded Green function are rewritten in terms  
$r^{(0)}$ and $\bm{r}$, we find that their reality
is preserved at all times.

Next we consider the Keldysh Green function. In order to calculate the 
time-evolution of the order parameter we only need to know the Keldysh
Green function at coinciding times. 
Here the parameterization 
\begin{equation}
K_m(t,t)=i\,\eta_3~\bm{k}_m(t)\cdot\bm{\eta},
\end{equation}
in terms of a real vector 
\begin{equation}
\bm{k}_m(t)=\left(k^{(1)}_m(t),k^{(2)}_m(t),k^{(3)}_m(t)\right),
\end{equation}
is respected by the initial condition and preserved by the equations
of motion. 

From the equations of motion (Eq.~\ref{Dyn:Dyson}) we derive differential equations
\begin{equation}
\frac{d}{dt}r_m(t,\tau)=b_m(t)r_m(t,\tau)-r_m(t,\tau)b_m(t-\tau),\label{Dyn:drdt}
\end{equation}
\begin{equation}
\frac{d}{dt}\bm{k}_m(t)+2\bm{b}_m(t)\times\bm{k}_m(t)+2E_{\rm Th}\bm{k}_m(t)=4 E_{\rm Th}\bm{f}_m(t),
\label{Dyn:dkdt}
\end{equation}
for the matrix $r_m(t,\tau)$ and the vector $\bm{k}_m(t)$.
The equation (\ref{Dyn:dkdt}) with $E_{\rm Th}=0$ was studied in Refs.~\onlinecite{Bar06,Yuz05}. 
In these references the dynamics of the order parameter of an isolated superconductor
was calculated. We see that coupling the system to leads introduces two terms proportional
to $E_{\rm Th}$. One (on the left-hand side of Eq.~(\ref{Dyn:dkdt})) can be considered
a damping term and is proportional to ${\bm k}_m(t)$. The other (on the right-hand
side of Eq.~(\ref{Dyn:dkdt})) can be considered a driving or source term.
  
In Eq.~(\ref{Dyn:dkdt}), $b_m(t)$ is a matrix
and $\bm{b}_m(t)$ a vector such that
\begin{subequations}
\begin{eqnarray}
b_m(t)&=&i\,\bm{b}_m(t)\cdot\bm{\eta},\\
\bm{b}_m(t)&=&\left({\rm Re}\,\Delta(t),-{\rm Im}\,\Delta(t),\mu_s(t)-\varepsilon_m\right).
\end{eqnarray}
\end{subequations}
The equation for $\bm{k}_m(t)$ contains a source term $4E_{\rm Th}\bm{f}_m(t)$. The vector 
$\bm{f}_m(t)$ is given by
\begin{widetext}
\begin{equation}
\bm{f}_m(t)=\int_0^\infty d\tau\,\left[ r_m^{(0)}(t,\tau)\bm{s}(t,\tau)-\bm{r}_m(t,\tau)s^{(0)}(t,\tau)
-\bm{r}_m(t,\tau)\times\bm{s}(t,\tau)\right].
\end{equation}
\end{widetext}
In this equation the scalar function $s^{(0)}(t,\tau)$ and the vector $\bm{s}(t,\tau)$ 
parameterize the Keldysh component of the self-energy as follows
\begin{subequations}
\begin{eqnarray}
\Sigma_K(t,t')&=&2E_{\rm Th}\,\eta_3~s(t,\tau),\\
s(t,\tau)&=&s^{(0)}(t,\tau)\,\eta_0+i\bm{s}(t,\tau)\cdot\bm{\eta}.
\end{eqnarray}
\end{subequations}
Referring back to Sec.~\ref{sec:model}, where $\Sigma_K$ is expressed in terms of the Fourier transform of
the reservoir filling factors, we find explicitly
\begin{subequations}
\begin{eqnarray}
s^{(0)}(t,\tau)&=&\frac{1}{\pi}\mathcal P\left(\frac{1}{\tau}\right)\cos\frac{\phi(t)-\phi(t-\tau)}{2},\\
\bm{s}(t,\tau)&=&-\frac{\gamma}{\pi\tau}\left(0,0,\sin\frac{\phi(t)-\phi(t-\tau)}{2}\right).\nonumber\\
\end{eqnarray}
\end{subequations}

By imposing self-consistency, the order parameter $\Delta(t)$ is expressed in terms of 
the components of $\bm{k}_m(t)$ as
\begin{equation}
\Delta(t)=\frac{g\delta_s}{2}\sum_{m=-\Omega}^\Omega k_m^{(1)}(t)-ik_m^{(2)}(t),\label{Dyn:selfcond}
\end{equation}
where the number of levels on the island is $2\Omega+1$.
This makes the differential equations non-linear, since they contain terms
in which $\Delta(t)$ multiplies ${\bm k}_m$ and $r$.
We eliminate the dimensionless pairing strength $g$ and the mean level spacing $\delta_s$ in favor of 
$\Delta_{\rm eq}$, the order parameter of the island in equilibrium, by means of the equilibrium
self-consistency relation
\begin{equation}
\frac{2}{g\delta_s}=\sum_{m=-\Omega}^\Omega\frac{1}{\xi_m}\frac{2}{\pi}\arctan\frac{\xi_m}{E_{\rm Th}},
\end{equation}
where
\begin{equation}
\xi_m=\sqrt{\varepsilon_m^2+\Delta_{\rm eq}^2}.
\end{equation}

As with $\Delta(t)$, the chemical potential $\mu_s(t)$ is determined by a self-consistency equation.
The chemical potential takes into account the work that must be performed against the electric
field of the excess charge on the superconductor in order to add more charge. 
Thus $\mu_s(t)$ is related to the charge of the island by
$\mu_s(t)=e[Q(t)-Q_0]/C$ where $C$ is the capacitance of the island. 
In this equation $Q_0$ represents the fixed positive background charge
and $Q(t)$ is the combined charge of all the electrons on the island. Since the differential
equations (\ref{Dyn:drdt}) and (\ref{Dyn:dkdt}) only depend on the difference $\mu_s(t)-\mu_s(t-\tau)$,
the positive background charge need not be specified. The charge $Q(t)$ is related to the Keldysh Green
function. Indeed, the average number $n_m(t)$ of electrons (with spin-degeneracy included) in level $m$
at time $t$ is given by $n_m(t)=(1-i{\rm Tr}[K_m(t,t)])/2$. Hence $\mu_s(t)$ is related to $\bm{k}_m$
by the equation
\begin{equation}
\mu_s(t)-\mu_s(t-\tau)=\frac{e^2}{2C}\sum_{m=-\Omega}^\Omega k_m^{(3)}(t)-k_m^{(3)}(t-\tau).\label{Dyn:selfconm}
\end{equation}

Finally, we have to specify the initial conditions for $r_m(t,\tau)$ and $\bm{k}_m(t)$.
We will assume for our simulations that 
the voltage between the reservoirs is zero
and the system is in zero-temperature equilibrium for times $t<0$.
The corresponding initial condition at $t=0$ is
\begin{subequations}
\begin{eqnarray}
r^{(0)}_m(0,\tau)&=&-\theta(\tau)e^{-E_{\rm Th}\tau}\cos(\xi_m\tau),\\
\bm{r}_m(0,\tau)&=&\theta(\tau)e^{-E_{\rm Th}\tau}\frac{\sin(\xi_m\tau)}{\xi_m}\left(-\Delta_{\rm eq},0,\varepsilon_m\right),\nonumber\\
\\
\bm{k}_m(0)&=&\frac{1}{\xi_m}\frac{2}{\pi}\arctan\frac{\xi_m}{E_{\rm Th}}\left(\Delta_{\rm eq},0,-\varepsilon_m\right).
\end{eqnarray}
\end{subequations}

\begin{figure}[tbh]
\begin{center}
\includegraphics[width=.99 \columnwidth]{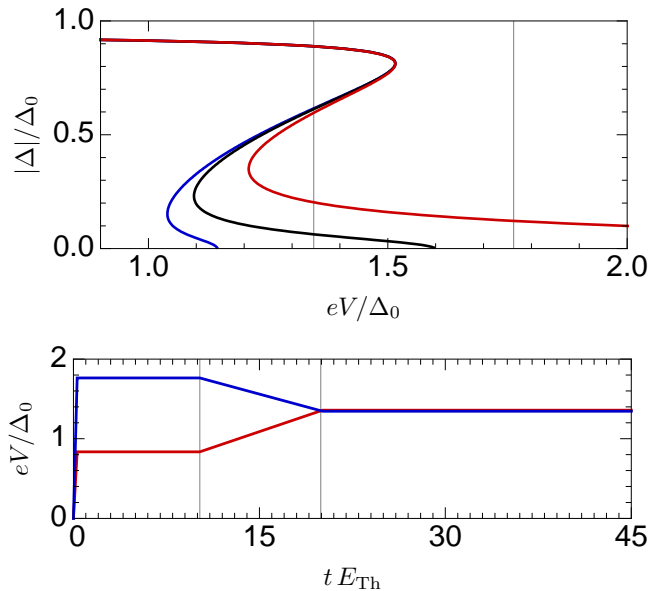}
\caption{Top panel: The stationary solutions for
$\Delta$ vs. the bias voltage $V$, corresponding to the parameters used in generating Fig.~\ref{figdeltat}.
The outer curve (red) corresponds to $E_{\rm Th}=0.069\Delta_0$ and $\gamma=0.2$.
The middle curve (black) corresponds to $E_{\rm Th}=0.069\Delta_0$ and $\gamma=0.1$.
The inner curve (blue) corresponds to $E_{\rm Th}=0.069\Delta_0$ and $\gamma=0.05$.
The vertical lines indicate $V_f$ and $V_2$. ($V_1$ is beyond the left edge of the figure.)
Bottom panel: The time-dependence of the voltage. The upper (blue) curve corresponds to the
blue curves of $\Delta$ vs. $t$ in Fig.~\ref{figdeltat}. The lower (red) curve corresponds to the
red curves of $\Delta$ vs. $t$ in Fig.~\ref{figdeltat}. 
\label{figv}} 
\end{center}
\end{figure}

We are now ready to study the time-evolution of $\Delta(t)$ when a non-zero bias voltage 
$V(t)$ between the leads is present for times $t>0$.
In the calculations we report on here, we worked with 
$E_{\rm Th}=0.069\,\Delta_0$ and three different $\gamma$, namely
$\gamma=0.05$, $\gamma=0.1$ and $\gamma=0.2$.
These all correspond to points from regions 
$C$ and $D$ in the $E_{\rm Th}$--$\gamma$
parameter space of Fig.~\ref{fig:parspace}.
Hence, for each of the parameter choices, there is a bias voltage interval 
$[V_-,V_+]$ where there are more than one
non-zero stationary solutions for $|\Delta|$.
The three curves of stationary $|\Delta|$ versus $V$, corresponding to the 
different parameter choices, are plotted in the top panel of Fig.~\ref{figv}.

For given $E_{\rm Th}$ and $\gamma$ we did two numerical runs with different
time-dependent voltages $V(t)$. 
The two voltages are plotted as functions of time 
in the bottom panel of Fig.~\ref{figv}. 
In the first run we start
by rapidly establishing a bias voltage $V_1<V_-$. 
Rapid here means $dV/dt\gg\Delta_0E_{\rm Th}/e$.
In this case $V$ changes by an amount of order $\Delta_0/e$ --- the scale at which the
stationary solution for $|\Delta|$ depends on $V$ --- in a time that is short compared to the
relaxation time $E_{\rm Th}^{-1}$. (Slow refers to the opposite limit.)
We then keep the voltage constant at $V_1$ for a length of time of several $E_{\rm Th}$.
This time-interval is long enough for any transient behavior induced by the rapid change of 
$V(t)$ to disappear.  We then slowly increase
the bias voltage until we reach a bias voltage $V_f\in[V_-,V_+]$
for which more than one non-zero stationary solutions exist.
In the second run we start by rapidly establishing a bias voltage $V_2>V_+$. 
We keep the voltage fixed at $V_2$ for a time of several $E_{\rm Th}^{-1}$.
We then slowly decrease the voltage to
$V_f$. The values of $V_1$, $V_2$ and $V_f$ were chosen
$V_1=0.83 \Delta_0$, $V_2=1.76 \Delta_0$ and $V_f=1.34\Delta_0$. 
The calculations were performed with $501$ equally spaced levels with level spacing 
$\delta_s=0.018\Delta_0$ and the capacitance was chosen $C=0.1\,e^2/\Delta_0$.

\begin{figure}[tbh]
\begin{center}
\includegraphics[width=.99 \columnwidth]{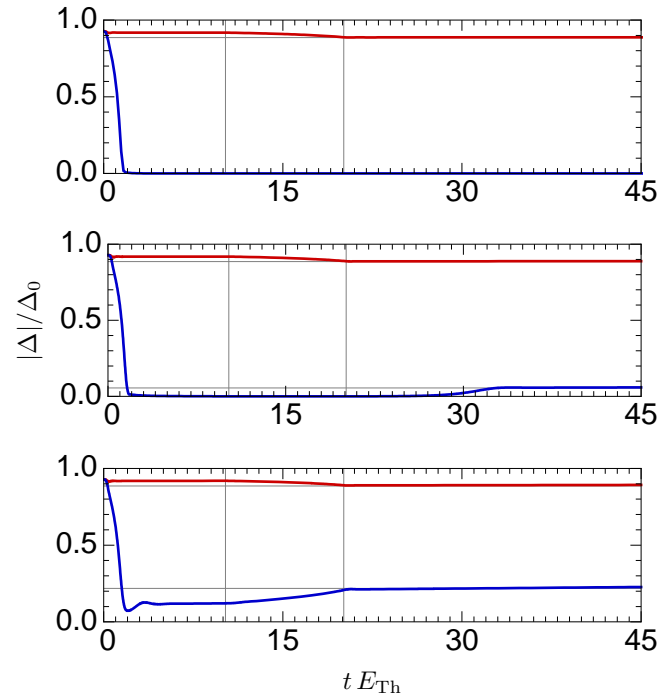}
\caption{The amplitude of the order parameter as a function of time. All curves are for $E_{\rm Th}=0.069\Delta_0$.
The top, middle and bottom panels correspond to $\gamma=0.05$, $\gamma=0.1$ and $\gamma=0.2$ respectively.
The red curves correspond to a voltage that is increased from $V_1=0.83\Delta_0$ to $V_f=1.34\Delta_0$. 
The blue curves 
correspond to the voltage being decreased from $V_2=1.76\Delta_0$ to $V_f=1.34\Delta_0$. The vertical lines indicate the
time-interval in which the voltage changes from either $V_1$ or $V_2$ to $V_f$.
The thin horizontal lines correspond to the stationary values of $|\Delta|$ for a bias voltage $V=V_f$
as calculated from Eq.~(\ref{eq:transcendental}). 
\label{figdeltat}}
\end{center}
\end{figure}

\begin{figure}[tbh]
\begin{center}
\includegraphics[width=.99 \columnwidth]{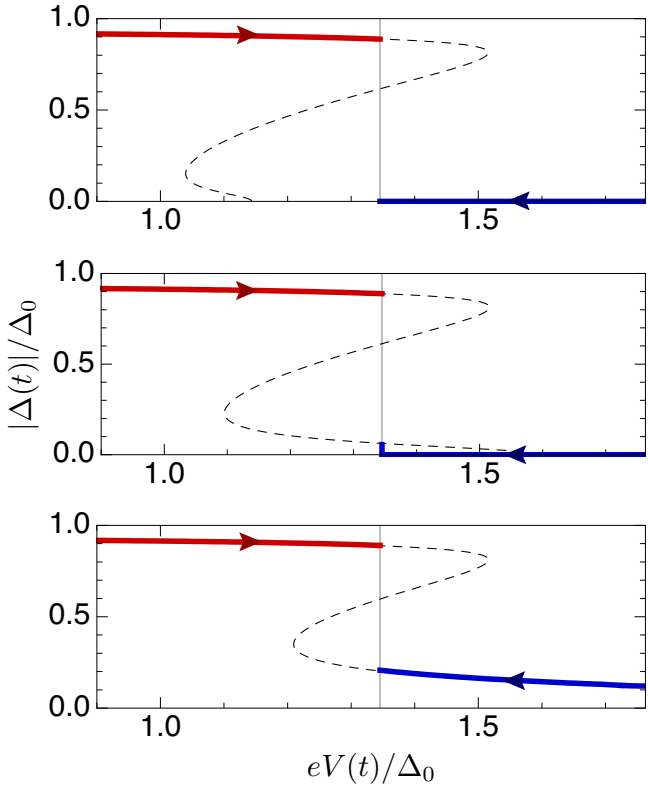}
\caption{The amplitude of the order parameter $|\Delta(t)|$ as a function of voltage $V(t)$. 
The parameter values of the three panels are the same as those in Fig.~\ref{figdeltat}, i.e.
all curves are for $E_{\rm Th}=0.069\Delta_0$.
The top, middle and bottom panels correspond to $\gamma=0.05$, $\gamma=0.1$ and $\gamma=0.2$ respectively.
The red curves correspond to a voltage that is increased from $V_1=0.83\Delta_0$ to $V_f=1.34\Delta_0$. 
The blue curves 
correspond to the voltage being decreased from $V_2=1.76\Delta_0$ to $V_f=1.34\Delta_0$. 
The dashed lines represents the stationary value of $|\Delta|$ vs. $V$, as calculated in 
Sec.~\ref{sec:stat} and plotted in Fig.~\ref{figv}. 
\label{figdv}}
\end{center}
\end{figure}

The resulting $|\Delta|$ are plotted as functions 
of time in Fig.~\ref{figdeltat}. 
They firstly show that after the initial
rapid change in the bias voltage
the system always relaxes into a stationary state consistent with the new voltage. The
relaxation takes a time of the order $E_{\rm Th}^{-1}$.
Secondly, if the system is in a stationary state, and the bias voltage is changed slowly 
then $|\Delta(t)|$ adiabatically
tracks the stationary solution corresponding to the instantaneous value of the voltage.
This is seen most clearly in Fig.~\ref{figdv} where we plot $|\Delta(t)|$ as a function of $V(t)$ 
and compare this to the stationary $|\Delta|$ vs. constant $V$ curves. 
Our prediction about hysteresis is confirmed. Systems with different histories end up in different
stationary states at the same voltage bias. If the voltage is slowly swept from a small initial
voltage to $V_f\in[V_-,V_+]$ a stationary state with a large value for $|\Delta|$ is reached. 
If the voltage is swept from a large initial voltage to $V_f\in[V_-,V_+]$, a stationary
state is reached that corresponds to a small value of $|\Delta|$. We must mention here that
we observe some slow drift (too slow to be visible in Fig.~\ref{figdeltat}) 
in $|\Delta|$  after  the voltage has reached $V_f$.
The value of $|\Delta|$ seems to increase linearly at a rate 
$d|\Delta|/dt\sim10^{-4}\Delta_0^2$. Within the numerical accuracy of
the calculation, this is negligible and we believe the drift is simply 
an artifact of the numerics. 

In our data there is one exception to the rule of adiabatic evolution.
In the middle panel of Fig.~\ref{figdeltat}, 
$|\Delta(t)|$ takes much longer that $E_{\rm Th}^{-1}$
to respond when the voltage is changed from $V_2$ to $V_f$. Hence, $|\Delta(t)|$ as a function
of $V(t)$ does not track the stationary solution in this instance.
The reason is the following:
For a voltage $V=V_2$, the only stationary solution has $\Delta=0$. 
When the voltage is decreased to $V_f$, a non-zero
stationary solution for $\Delta$ exists. However $\Delta=0$ is still a valid state,
albeit unstable. The time it takes the system to diverges from the unstable state is not
determined by $E_{\rm Th}$ but rather by small numerical errors that perturb the unstable state.


One possible explanation for the observed stability of the stationary states is overdamping.
According to this hypothesis, if we decrease the Thouless energy further, thereby decreasing
the damping, the stationary solutions will become unstable. Some evidence for the hypothesis
might be visible in Fig.~\ref{figdeltat}. After the voltage is changed rapidly, we might 
expect $|\Delta(t)|$ to perform damped oscillations while relaxing to the new stationary 
state. However in Fig.~\ref{figdeltat} no such oscillations are visible, apparently
implying that the relaxation rate is larger than the oscillation frequency. There
is however another possible explanation for the lack of oscillatory behavior after an abrupt change
in $V$. The argument is that an abrupt change in $V$ cannot be communicated to the
system abruptly, but only at a rate comparable to the damping rate $E_{\rm Th}$. This
is because the superconductor learns of the change in voltage by the same mechanism as 
by which damping occurs, that is, by tunneling of particles between the leads and the island.
Hence the response of the order parameter is always gradual. 

How do we test whether overdamping hypothesis is true or false? Ideally we would have
liked to repeat the above numerical calculation with a smaller value of $E_{\rm Th}$
and see if the stationary states are still stable.   
However, the value $E_{\rm Th}=0.069\Delta_0$ that we used above is close to 
the smallest value for which we can do reliable numerics in reasonable time.
Since we cannot make $E_{\rm Th}$ smaller, we resolve the issue of overdamping as follows.
We compare the dynamics of $\Delta$
after an abrupt change in the pairing interaction strength $g$ at $E_{\rm Th}=0.069\Delta_0$ 
to the dynamics after a change in $g$ at $E_{\rm Th}=0$.\cite{Note4}
We know that in the isolated system, ($E_{\rm Th}=0$) $|\Delta|$ will perform persistent 
oscillations.\cite{Bar06,Yuz05} 
The period of oscillation gives a typical time-scale for the internal dynamics of $\Delta$.
If, in the open system (i.e. $E_{\rm Th}\not=0$), we observe a few damped oscillations (the more the better)
in $|\Delta|$ before the system relaxes to equilibrium,
it means that damping occurs at a timescale larger than that of the internal dynamics of
the superconductor. In this case the hypothesis of overdamping is discredited. 
  
\begin{figure}[tbh]
\begin{center}
\includegraphics[width=.99 \columnwidth]{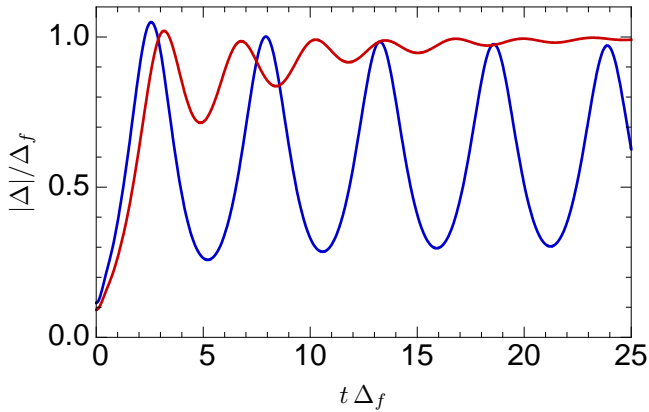}
\caption{The order parameter versus time after the pairing strength was increased from $\Delta_i=0.05\Delta_f$
to $\Delta_f$ abruptly at $t=0$. The blue curve is for an isolated superconductor while the
red curve is for a superconductor connected to leads. For this case a Thouless energy 
$E_{\rm Th}=0.075\Delta_f$ was used. The data was obtained using $501$ equally spaced levels with 
level spacing $\delta_s=0.02\Delta_f$. The capacitance was chosen $C=0.1e^2/\Delta_f$.\label{figchanged}}
\end{center}
\end{figure}

In our numerical implementation of the above,
we work with the following parameters: The initial pairing interaction is 
such that for $t<0$, $\Delta=\Delta_i$. The increased pairing interaction strength corresponds to
an equilibrium value of the order parameter $\Delta_f=20\Delta_i$.
The persistent oscillations of $|\Delta(t)|$ in the isolated system
are shown in the blue curve in Fig.~\ref{figchanged}. We repeat the calculation, now for a
superconductor connected to leads. We use a s energy $E_{\rm Th}=0.075\Delta_f$.
The result for $|\Delta(t)|$ in the presence of leads is the red curve in Fig.~\ref{figchanged}. 
We see that $|\Delta(t)|$
eventually decays to a constant, as expected.
The extent of the damping is such that several oscillations are completed within the decay time.
Hence we conclude that the numerical results that we obtained previously are outside the
regime of overdamping. It follows that the lack of oscillatory behavior in Fig.~\ref{figdeltat}
is due to the fact that the superconductor only gradually becomes aware of a change in the 
voltage. 

\section{Conclusion}
\label{sec:con}
We have studied a voltage biased NISIN junction i.e. a superconducting island
connected to normal leads by means of tunnel junctions.
We restricted ourselves to the regime where the dominant energy relaxation
mechanism in the superconductor is the tunneling of electrons from the
superconductor to the leads. We also restricted ourselves to the regime
of low transparency junctions where the position dependence of the order parameter
inside the superconductor can be neglected. 

In Sec.~\ref{sec:stat} we found the
stationary states of the system. For these, the order parameter $\Delta$ and
the chemical potential are implicitly determined by Eq.~(\ref{eq:transcendental}).
We also found the current between the leads [cf. Eq.~(\ref{eq:current})]. 
The most striking feature of the stationary states is that there can be
more than one stationary state at a given voltage. These are characterized by
different values of $|\Delta|$ and of the current as can be seen in the 
$I$-$V$ curves of Fig.~\ref{fig:current}. Depending on system parameters,
superconductivity can survive up to voltages large compared to $\Delta_0$, the
order parameter of the isolated superconductor. In this case, increasing the
voltage eventually leads to a second order phase transition to the normal state.
We have found that the critical voltage at which the transition occurs obeys a
power-law [cf. Eq.~(\ref{eq:power law})]. 

In Sec.~\ref{sec:dyn} we studied time-dependent states of the system. In this way
we were able to demonstrate the stability of the stationary states we have found
in the previous section. Our results also indicate that a DC biased system always
relaxes into a stationary state. In the parameter region of multiple stationary 
states we demonstrated bi-stability. Associated with this are first order 
phase-transitions: there are critical voltages where $\Delta$ (and the current)
make finite jumps. Furthermore, there is hysteresis of $|\Delta|$ and the current associated with
the bi-stability.

\acknowledgments
This research was supported by the Dutch Science Foundation NWO/FOM.

\end{document}